%% file: paper_1250.tex
\RequirePackage[left]{lineno}
\documentclass[aps,prl,twocolumn,showpacs,preprintnumbers,amsmath,amssymb,refmerge,superscriptaddress]{revtex4}
\usepackage{color}
\usepackage{graphicx}
\usepackage{dcolumn}
\usepackage{longtable}
\usepackage[percent]{overpic}

\def\babar  {\mbox{\slshape B\kern-0.1em{\smaller A}\kern-0.1em B\kern-0.1em{\smaller A\kern-0.2em R}}}

\newcommand{\BSGG}{\mbox{$B_{s}^{0}\rightarrow\gamma\gamma$}}
\newcommand{\BSPG}{\mbox{$B_{s}^{0}\rightarrow\phi\gamma$}}

\newcommand{\BSTBST}{\mbox{$B_{s}^{*}\bar B_{s}^{*}$}}
\newcommand{\BSTBS}{\mbox{$B_{s}^{*}\bar B_{s}$}}
\newcommand{\BSBS}{\mbox{$B_{s}\bar B_{s}$}}
\newcommand{\BKG}{\mbox{$B^{0}\rightarrow K^{*}(892)^{0}\gamma$}}

\newcommand{\BXsg}{\mbox{$B\rightarrow X_{s}\gamma$}}

\newcommand{\bs}{\begin{sideways}}
\newcommand{\es}{\end{sideways}}
\newcommand{\bc}{\begin{center}}
\newcommand{\ec}{\end{center}}

\makeatletter
\newcommand*{\rom}[1]{\expandafter\@slowromancap\romannumeral #1@} 
\makeatother

\begin{document}

\preprint{\vbox{
                 \hbox{\bf Belle Preprint 2014-18}
                 \hbox{\bf KEK Preprint 2014-32}                  
                 \hbox{}
}}

\title{\textbf{Search for $\boldmath{\BSGG}$ and a measurement of the branching fraction for $\boldmath{\BSPG}$}}


\input{author_pub435}


\begin{abstract}
We search for the decay $\BSGG$ and measure the branching fraction for $\BSPG$ using 121.4~$\textrm{fb}^{-1}$ of data collected at the $\Upsilon(\mathrm{5}S)$ resonance with the Belle detector at the KEKB asymmetric-energy $e^{+}e^{-}$ collider. The $\BSPG$ branching fraction is measured to be $(3.6 \pm 0.5 (\mathrm{stat.}) \pm 0.3 (\mathrm{syst.}) \pm 0.6 (f_{s})) \times 10^{-5}$, where $f_{s}$ is the fraction of $B_{s}^{(*)}\bar{B}_{s}^{(*)}$ in $b\bar{b}$ events. Our result is in good agreement with the theoretical predictions as well as with a recent measurement from LHCb. We observe no statistically significant signal for the decay $\BSGG$ and set a $90\%$ confidence-level upper limit on its branching fraction at $ 3.1 \times 10^{-6}$. This constitutes a significant improvement over the previous result.
\end{abstract}


\maketitle





\begin{figure}[hbp]
\centering
\includegraphics[scale=0.195]{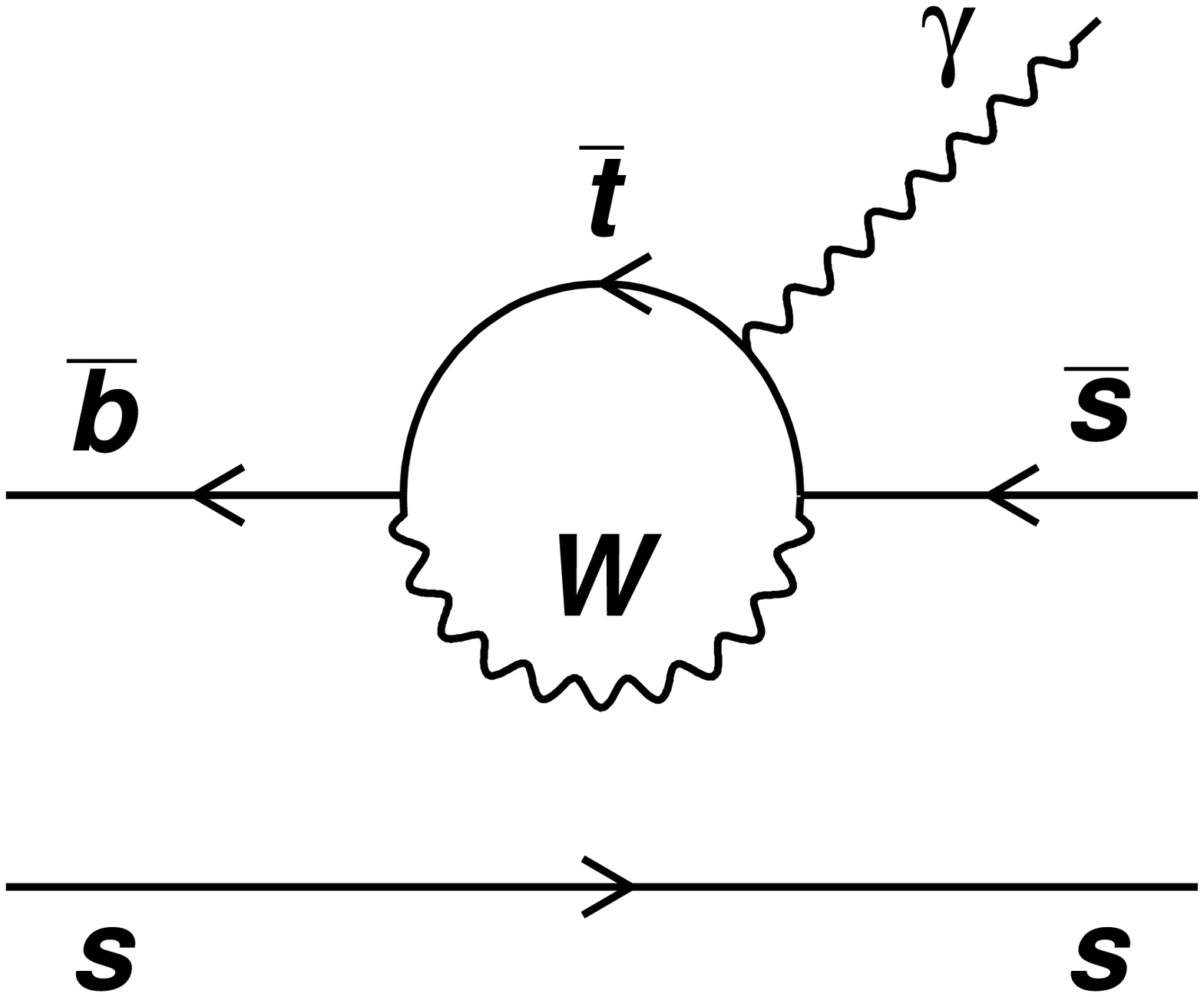}
\includegraphics[scale=0.195]{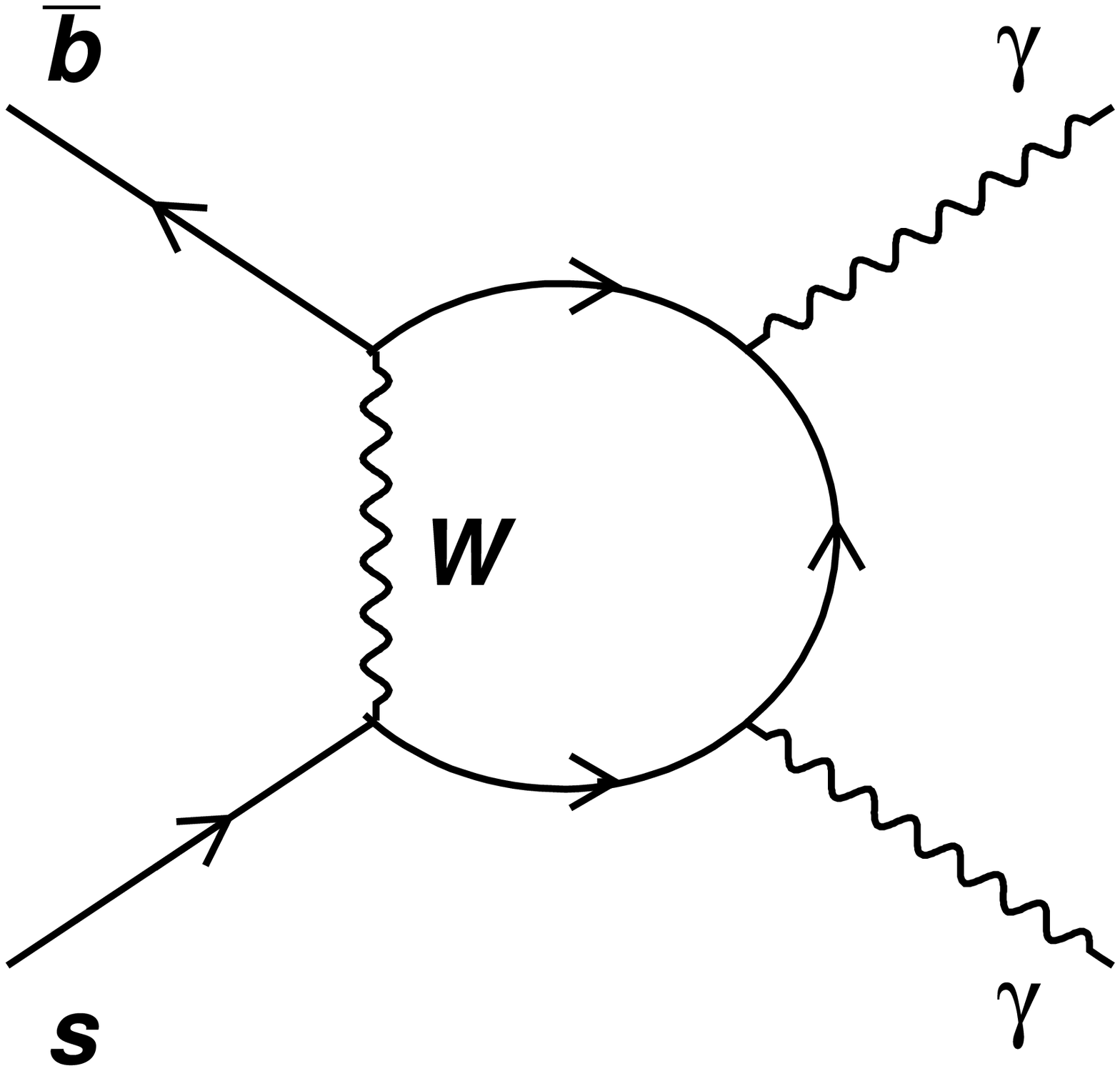}
\centerline{\hfill (a) \hfill \hfill (b) \hfill}
\smallskip
\caption{Leading-order Feynman diagrams for the decays (a) $\BSPG$ and (b) $\BSGG$.}
\medskip
\label{fig:Feynman}
\end{figure}

In the Standard Model (SM), the exclusive decays $\BSGG$ and $\BSPG$ are explained by the radiative transitions $b\rightarrow s\gamma\gamma$ and $b\rightarrow s\gamma$, respectively. The leading-order Feynman diagrams for these processes are shown in Fig.~\ref{fig:Feynman}. Within the SM framework, the branching fraction (BF) for $\BSPG$ is expected to be about $4\times 10^{-5}$ with $30\%$ uncertainty~\cite{Greub_Bspg, Ball_Bspg}. First observation of this decay was made by the Belle Collaboration using 23.6 $\textrm{fb}^{-1}$ of data collected at the $\Upsilon(5S)$ resonance and its BF was measured to be $(5.7^{+2.2}_{-1.9})\times 10^{-5}$~\cite{Jean_paper}. The latest tabulated world-average value is $(3.6 \pm 0.4) \times 10^{-5}$~\cite{PDG}. These experimental results are in good agreement with the theoretical expectations. Furthermore, the good agreement between theory and experimental results on exclusive decays mediated by $b\rightarrow s\gamma$ transitions~\cite{Greub_Bspg,Ball_Bspg,BFkg,HFAG} as well as on inclusive $\BXsg$ rates~\cite{B-Xsg_th1, B-Xsg_th2, HFAG} rules out large contributions to $\BSPG$ from physics beyond the SM. However, potential contributions from new physics could remain hidden within the large uncertainties in the SM predictions~\cite{RPV,Extradim}. The decay $\BSGG$ has not been observed yet. Currently, the upper limit on its BF is $8.7\times10^{-6}$ at 90$\%$ confidence level (CL)~\cite{Jean_paper}. This is almost an order of magnitude larger than the range covered by published theoretical calculations~\cite{Reina_Bsgg, Buchalla_Bsgg, Giri}. The $\BSGG$ BF is also constrained by the $\BXsg$ results in the $R$-parity conserving SUSY scenario~\cite{RPV}. However, in the $R$-parity violating (RPV) case, the possible contribution from $\lambda$-irreducible diagrams~\cite{lambda_irre_diag} (which have a negligible impact on the $b\rightarrow s\gamma$ amplitude at one loop) may enhance its BF by more than an order of magnitude~\cite{RPV}.

The results presented in this paper are based on 121.4~$\textrm{fb}^{-1}$ of data collected at the $\Upsilon(5S)$ resonance with the Belle detector~\cite{detector,detector2} at the KEKB~\cite{accelerator} asymmetric-energy B-factory at KEK in Japan. The Belle detector consists of a 4-layer silicon vertex detector (SVD), a central drift chamber (CDC), aerogel Cherenkov counters (ACC), time-of-flight counters (TOF) and an electromagnetic calorimeter (ECL). These detector components are located inside a solenoid with a magnetic field of 1.5~T whose flux-return yoke is instrumented to detect $K_{L}^{0}$ mesons and muons.


The $b\bar{b}$ production cross section at the $\Upsilon(5S)$ center of mass (CM) energy is measured to be $\sigma_{b\bar{b}}^{\Upsilon(5S)} = (0.340 \pm 0.016)$ nb~\cite{Sevda}, while the fraction of $B_{s}^{(*)}\bar{B}_{s}^{(*)}$ in $b\bar{b}$ events is $f_{s} = (17.2 \pm 3.0)\%$~\cite{Sevda}. The $B_{s}^{(*)}\bar{B}_{s}^{(*)}$ pairs include $\BSTBST$, $\BSTBS$ and $\BSBS$ with measured percentages $f_{\tiny{\BSTBST}} = (87.0 \pm 1.7)\%$ and $f_{\tiny{\BSTBS}} = (7.3 \pm 1.4)\%$~\cite{Sevda}. The $B_{s}^{*0}$ mesons decay to ground-state $B_{s}^{0}$ mesons through the emission of a photon. Charge conjugate modes are implied throughout this paper.


Signal Monte Carlo (MC) events for the decays, $\BSGG$ and $\BSPG$ are generated using {\textsc{EvtGen}}~\cite{EvtGen}; the response of the detector is simulated using {\textsc GEANT3}~\cite{GEANT}, with beam-related backgrounds from data added to the simulated samples. Charged tracks are required to originate from the interaction point (IP) by satisfying the criteria $\it{dr} <$ 0.5 cm and $\it{|dz|} <$ 3 cm, where $\it{|dz|}$ and $\it{dr}$ are the distances of closest approach to the IP along the $z$ axis (collinear with the positron beam) and in the transverse $r$-$\phi$ plane, respectively. Kaons are identified with an efficiency of about 85$\%$ by requiring $\mathcal{L}_{K}/(\mathcal{L}_{K}+\mathcal{L}_{\pi}) > 0.6$, where $\mathcal{L}_{K}$ and $\mathcal{L}_{\pi}$ are the likelihoods of the track being due to a kaon and pion, respectively, obtained using information from ACC, CDC and TOF. Tracks failing this requirement are assumed to be pions. To be reconstructed as a $\phi$ meson candidate, a pair of oppositely charged kaons must have an invariant mass within $\pm$12 MeV/$c^{2}$ ($\pm$ 2.5 $\sigma$) of the nominal $\phi$ mass. Similarly, the $K^{*0}$ candidates in the $\BKG$ control sample are reconstructed with oppositely charged kaon and pion candidates by requiring $|M_{K\pi} - m_{K^{*0}}| <$ 75 MeV/$c^{2}$, where $M_{K\pi}$ and $m_{K^{*0}}$ are the invariant mass of the kaon-pion pair and the nominal $K^{*0}$ mass, respectively. Photons are reconstructed by identifying energy deposits in the ECL not matched to any charged track and are required to have  a minimum energy of 100 MeV. To reject merged $\pi^{0}$ mesons and other neutral hadrons, the ratio of the energy deposited by a photon candidate in the $(3 \times 3)$ and $(5 \times 5)$ ECL crystal array centered on the crystal with the highest energy deposition is required to exceed 0.95. In the $B_{s}\rightarrow\gamma\gamma$ analysis, to reduce the effect of beam-related backgrounds, we use photons only from the barrel region ($\mathrm{33^{\circ} < \theta < 128^{\circ}}$, $\theta$ being the lab-frame polar angle). Daughter photons from $\pi^{0}$ and $\eta$ decays contribute to backgrounds for both $\BSPG$ and $\BSGG$. These are suppressed by applying a likelihood requirement based on the energies and polar angles of the photons and the diphoton invariant mass, calculated by combining the candidate photon with each other photon in the event. In addition, the timing characteristics of the energy clusters used for photon reconstruction are required to be consistent with the beam
collision time that is determined at the trigger level for the candidate event. To be considered as a $\BSGG$ ($\BSPG$) candidate, a pair of photons (a $\phi$ meson and a photon) needs to satisfy the requirements on the beam-energy constrained mass $M_{\rm bc}$ and energy difference $\Delta E$. These are defined as $M_{\rm bc} = \sqrt{(E_{\textrm{beam}}^{\textrm{CM}})^{2} - (p_{B_{s}}^{\textrm{CM}})^{2}}$ and $\Delta E $ = $E_{B_{s}}^{\textrm{CM}} - E_{\textrm{beam}}^{\textrm{CM}}$, where $E_{\textrm{beam}}^{\textrm{CM}}$ is the beam energy, and $p_{B_{s}^{0}}^{\textrm{CM}}$ and $E_{B_{s}^{0}}^{\textrm{CM}}$ are the momentum and energy, respectively, of the $B_{s}^{0}$ meson candidate, with all variables evaluated at the CM frame. Signal candidates are required to satisfy $M_{\rm bc} > 5.3$ GeV/$c^{2}$ for each mode, $-$0.4 GeV $< \Delta E < 0.1$ GeV for the $\BSPG$ mode and $-$0.7 GeV $< \Delta E < 0.2$ GeV for the $\BSGG$ mode. No events with multiple $B_{s}^{0}$ candidates are found in the signal MC sample, while the rate of multiple $B_{s}^{0}$ candidates in data is far below 1$\%$ for each analysis. Multiple candidates are removed by selecting the one with the more energetic photons.


The dominant source of background for both decay modes is the production of light quark-antiquark pairs ($q = u, d, s, c$) in the $e^{+}e^{-}$ annihilation, identified hereinafter as continuum. Since the quarks carry significant momenta, continuum events are jet-like and are therefore topologically different from isotropic $B_{s}^{(*)}\bar{B}_{s}^{(*)}$ events, where $B_{s}$ mesons carry much smaller momenta. To suppress this background, event shape variables such as the modified Fox-Wolfram moments~\cite{FoxWolfram} and the absolute value of the cosine of the angle between the thrust axis of the decay products of the $B_{s}$ candidate and the rest of the event are used as inputs to a Neural Network (NN)~\cite{NN}. The NN output ($\cal{C}_{\mathrm{NB}}$) is designed to peak at 1 for signal-like events and at $-$1 for background-like events. The NN output is also included in the unbinned maximum likelihood fit to extract the $\BSPG$ signal yield. As $\cal{C}_{\mathrm{NB}}$ peaks sharply at 1 and $-$1, it is very difficult to model it with a simple analytic function. Therefore, to improve the modeling, after rejecting the events with $\cal{C}_{\mathrm{NB}} <$ $\cal{C}_{\mathrm{NB_{min}}}$, a modified NN output is calculated as
\begin{linenomath}
\begin{equation}
\cal{C}'_{\mathrm{NB}} = \log \bigg(\frac{\cal{C}_{\mathrm{NB}} - \cal{C}_{\mathrm{NB_{min}}}}{\cal{C}_{\mathrm{NB_{max}}} - \cal{C}_{\mathrm{NB}}}\bigg),
\end{equation}
\end{linenomath}
where $\cal{C}_{\mathrm{NB_{min}}}$ =$-0.6$ and $\cal{C}_{\mathrm{NB_{max}}} \sim$ 1 are the lower and upper limits of $\cal{C}_{\mathrm{NB}}$ for the events used in the fit. For $\BSGG$, an optimized criterion of $\cal{C}_{\mathrm{NB}} >$ 0.77 is applied and this variable is excluded from the fit since considerable correlations are observed between $\cal{C_{\mathrm{NB}}}$ with each of the variables $M_{\rm bc}$ and $\Delta E$.


We perform a four-dimensional (two-dimensional) unbinned extended maximum likelihood fit comprising $M_{\rm bc}$, $\Delta E$, $\cos\theta_{\rm hel}$ and $\cal{C'_{\mathrm{NB}}}$ ($M_{\rm bc}$ and $\Delta E$) to extract the $\BSPG$ ($\BSGG$) signal yields. The $\phi$ helicity angle ($\theta_{\rm hel}$) is the angle between the $B_{s}^{0}$ momentum and that of one of the $\phi$ daughters in the $\phi$ rest frame. The total fit probability distribution function (PDF) consists of two components: signal and $q\bar{q}$ background. The signal component is further composed of signal coming from $\BSTBST$, $\BSTBS$ and $\BSBS$ decays, the relative fractions being fixed to the values measured in Ref.~\cite{Sevda}. Backgrounds arising from $B_{s}$ and non-$B_{s}$ decays are combined with the $q\bar{q}$ continuum as they have a small contribution and do not peak in the signal region. MC samples are used to parameterize the signal and background PDFs. The PDF for each component is represented by the product of one-dimensional functions since the correlations among the variables are negligible. The $M_{\rm bc}$, $\Delta E$, $\cos\theta_{\rm hel}$ and $\cal{C'_{\mathrm{NB}}}$ shapes of the $\BSPG$ signal are modeled with the sum of a Crystal Ball (CB)~\cite{CBall} and Gaussian function with a common mean, a CB function, a $\sin^{2}\theta_{\rm hel}$ distribution and the sum of two Gaussian functions, respectively. The background PDFs are described by an ARGUS function~\cite{Argus} for $M_{\rm bc}$ with its endpoint fixed to 5.434 GeV/$c^{2}$, a first-order Chebychev polynomial for $\Delta E$, a parabola for $\cos\theta_{\rm hel}$ and a Gaussian function for $\cal{C'_{\mathrm{NB}}}$. For the $\BSGG$ mode, the signal $M_{\rm bc}$ distributions are parameterized with a combination of CB and Gaussian functions with a common mean and the signal $\Delta E$ distributions are modeled with CB functions. The background is described by an ARGUS function for $M_{\rm bc}$ and a first-order Chebychev polynomial for $\Delta E$. For both analyses, the signal parameters are determined from MC except for the means and widths of the $M_{\rm bc}$ and $\Delta E$ distributions describing the $B_{s}^{*}\bar{B}_{s}^{*}$ contribution. The widths of $M_{\rm bc}$ and $\Delta E$ are calibrated using correction factors obtained from the $\BKG$ control sample. As a cross-check, the branching fraction of this mode is measured and found to be in good agreement with the world average~\cite{HFAG}. The $M_{\rm bc}$ mean is similarly adjusted using information from the $B_{s}\rightarrow D_{s} \pi$ analysis~\cite{Sevda}. The $\Delta E$ mean of the $\BSTBST$ component is allowed to float in the $\BSPG$ analysis. For the $\BSGG$ mode, we fix the $\Delta E$ mean to the signal MC value, as the correction to the $\Delta E$  mean, obtained from the $\BSPG$ analysis, is found to be within the statistical error. The uncertainty associated with this procedure is included as systematic uncertainty for this mode. All background parameters apart from the ARGUS endpoint are allowed to float. In total, we have nine free parameters for the $\BSPG$ fit and four free parameters for the $\BSGG$ fit.


In all three signal regions, we observe $91^{+14}_{-13}$ $\BSPG$ signal events with a significance of 10.7 $\sigma$, that includes the systematic uncertainties. The signal significance is evaluated as $\sqrt{-2\ln(\mathcal{L}_{0}/\mathcal{L}_{\mathrm{max}})}$, where $\mathcal{L}_{0}$ and $\mathcal{L}_{\mathrm{max}}$ are the likelihood values when the signal yield is constrained to 0 and when it is optimized, respectively. Systematic uncertainties are included by convolving the likelihood curve with a Gaussian function of width equal to the additive systematics.

\begin{figure}[htbp]
\centering
\begin{tabular}{cccc}
\label{fig:a}\includegraphics[width=0.25\textwidth,height=0.18\textwidth]{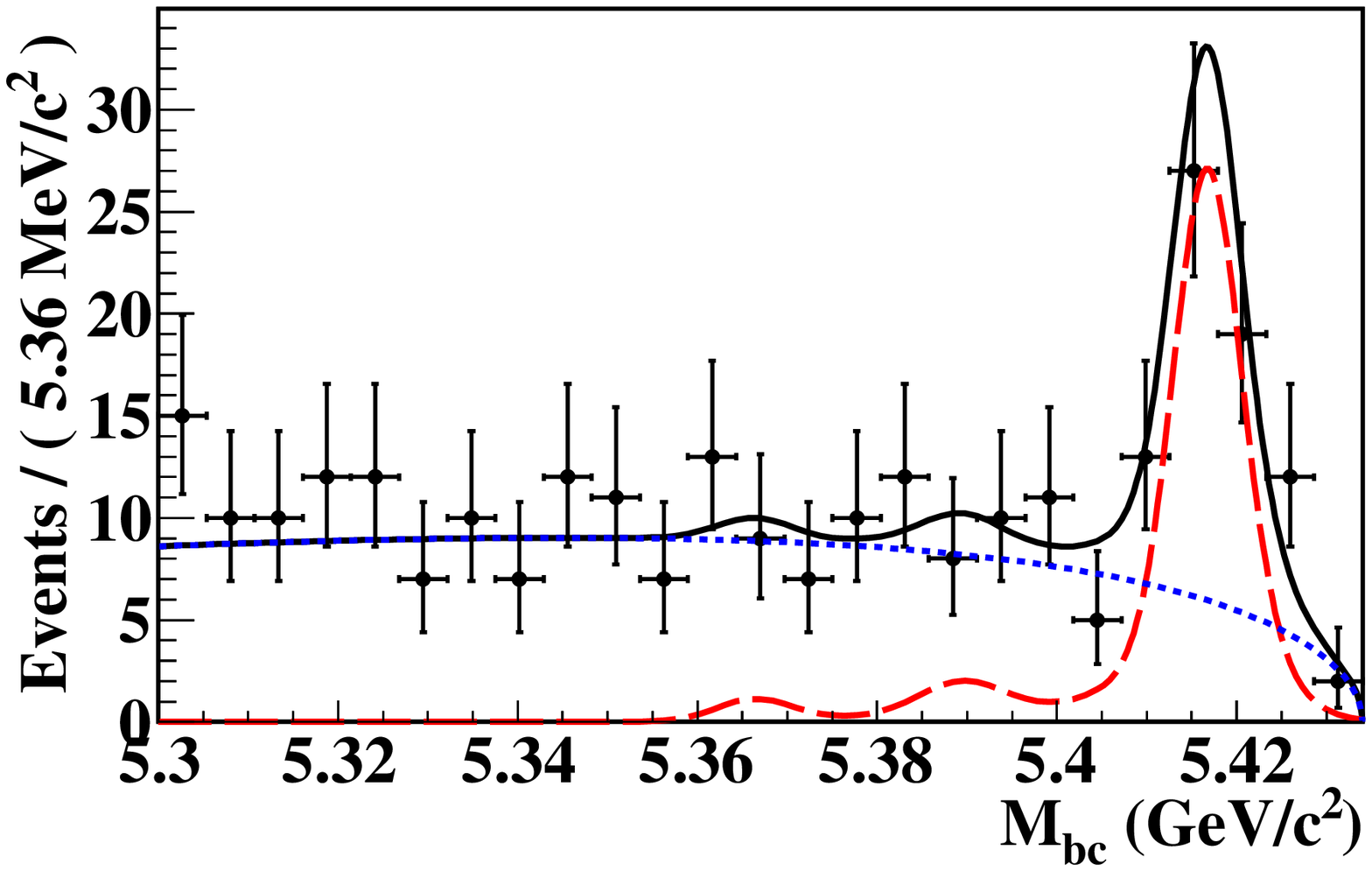}&
\label{fig:b}\includegraphics[width=0.25\textwidth,height=0.18\textwidth]{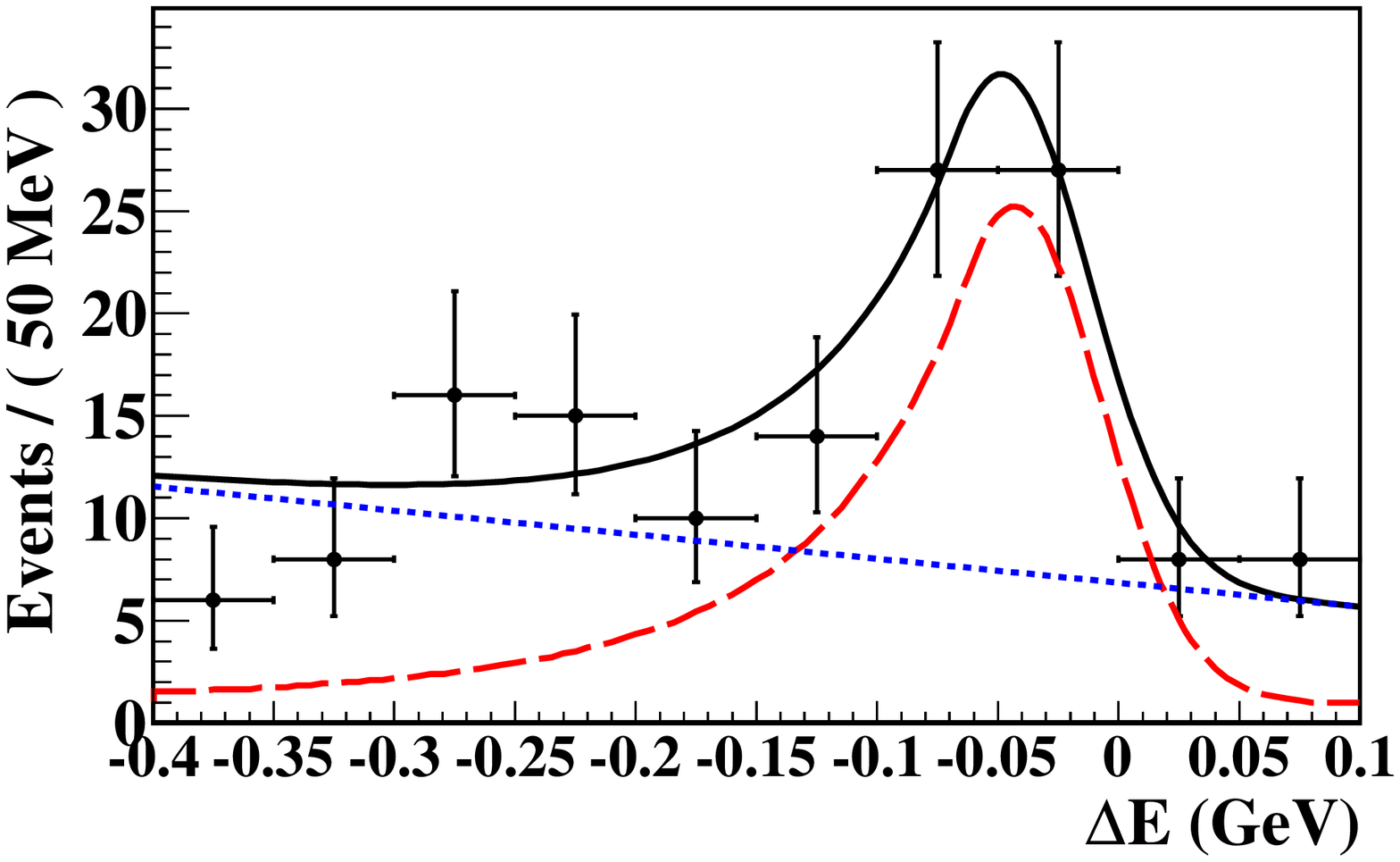}\\
\label{fig:c}\includegraphics[width=0.25\textwidth,height=0.18\textwidth]{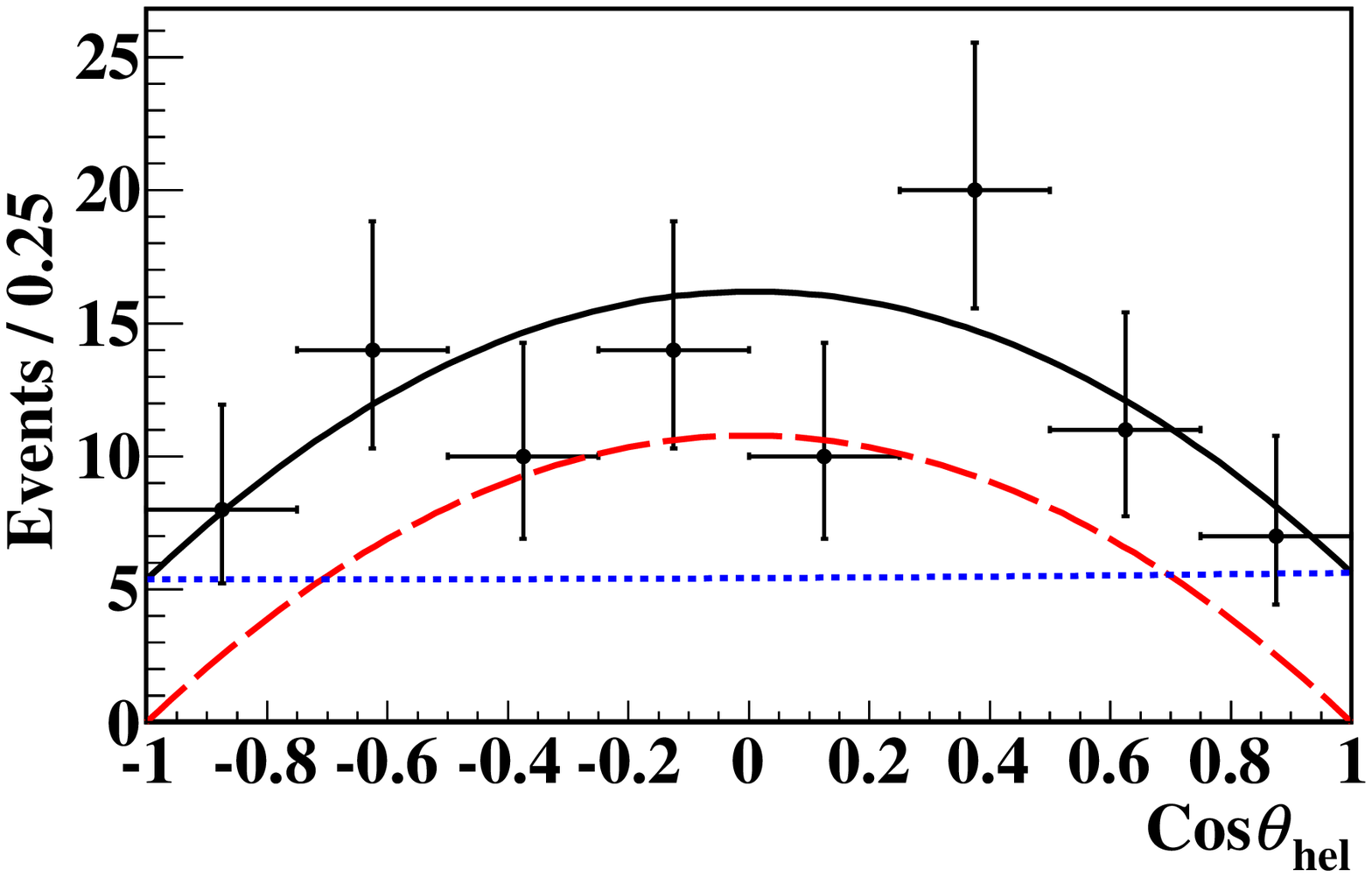}&
\label{fig:d}\includegraphics[width=0.25\textwidth,height=0.18\textwidth]{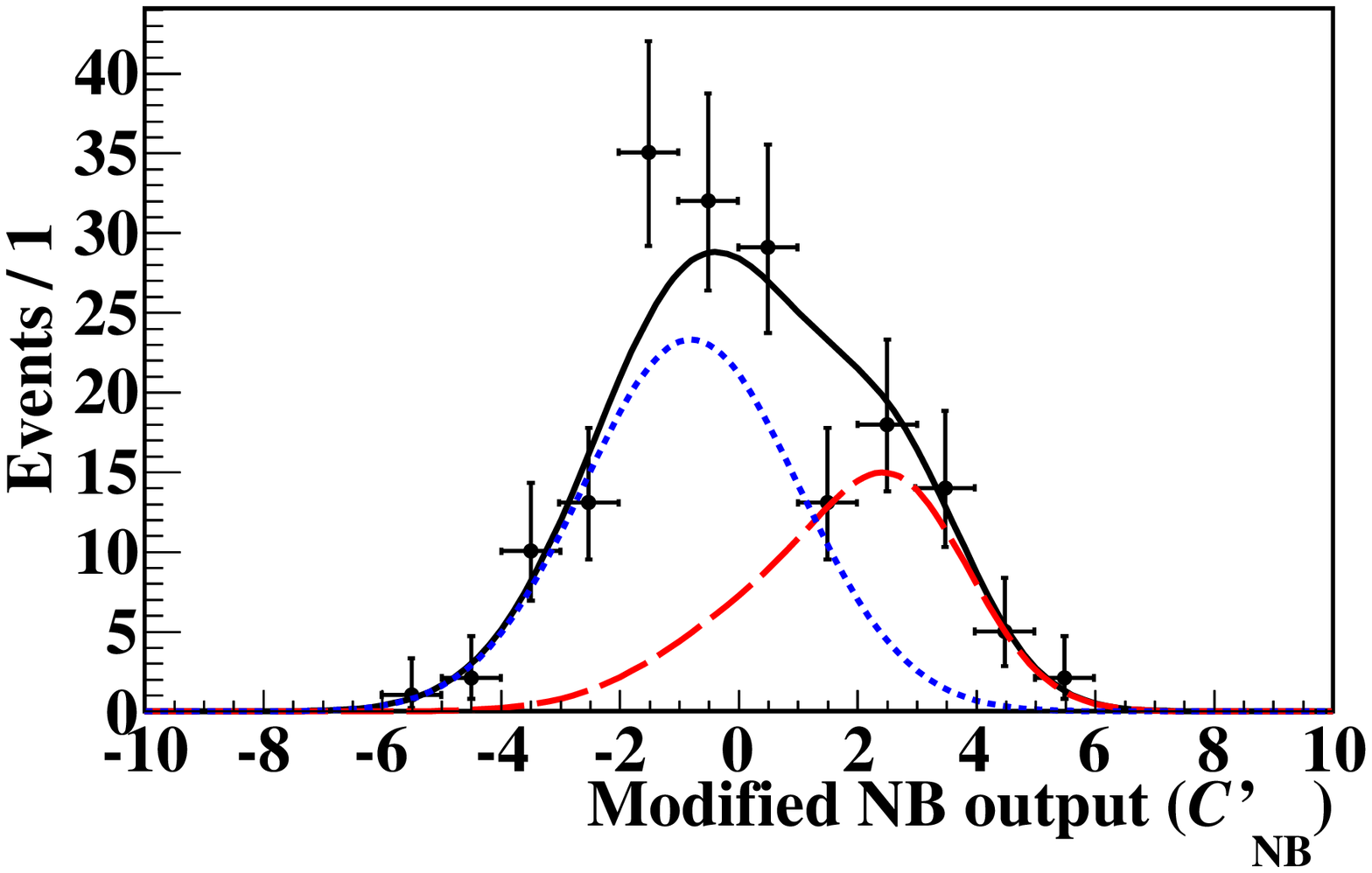}\\
\end{tabular}
\caption{Data fits for the $\BSPG$ analysis. The projections are shown only for events inside the $\BSTBST$ signal region except for the plotted variable. The $\BSTBST$ signal region is defined as $M_{\rm bc} > 5.4$ GeV/$c^{2}$, $-$0.2  $< \Delta E <$ 0.02 GeV, $|\cos\theta_{\rm hel}| <$ 0.8 \textrm{and} 0.0 $< \cal{C}'_{\mathrm{NB}} <$ 10.0. The points with error bars represent the data, the solid black curve represents the total fit function, the red dashed (blue dotted) curve represents the signal (continuum background) contribution.}
\label{fig:fit_pg}
\end{figure}

\begin{figure}[htbp]
\centering
\begin{tabular}{cc}
\includegraphics[width=0.25\textwidth,height=0.18\textwidth]{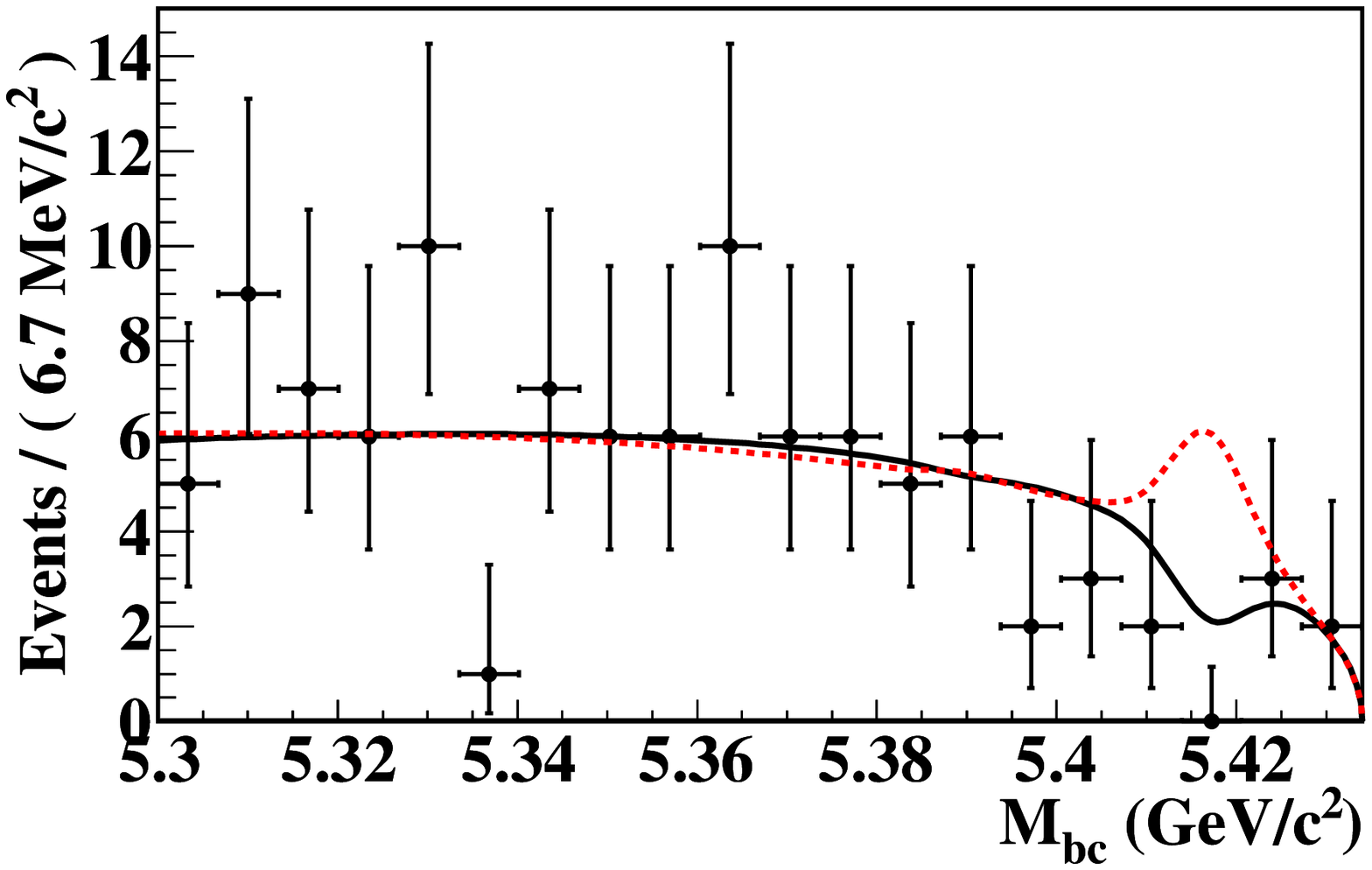}&
\includegraphics[width=0.25\textwidth,height=0.18\textwidth]{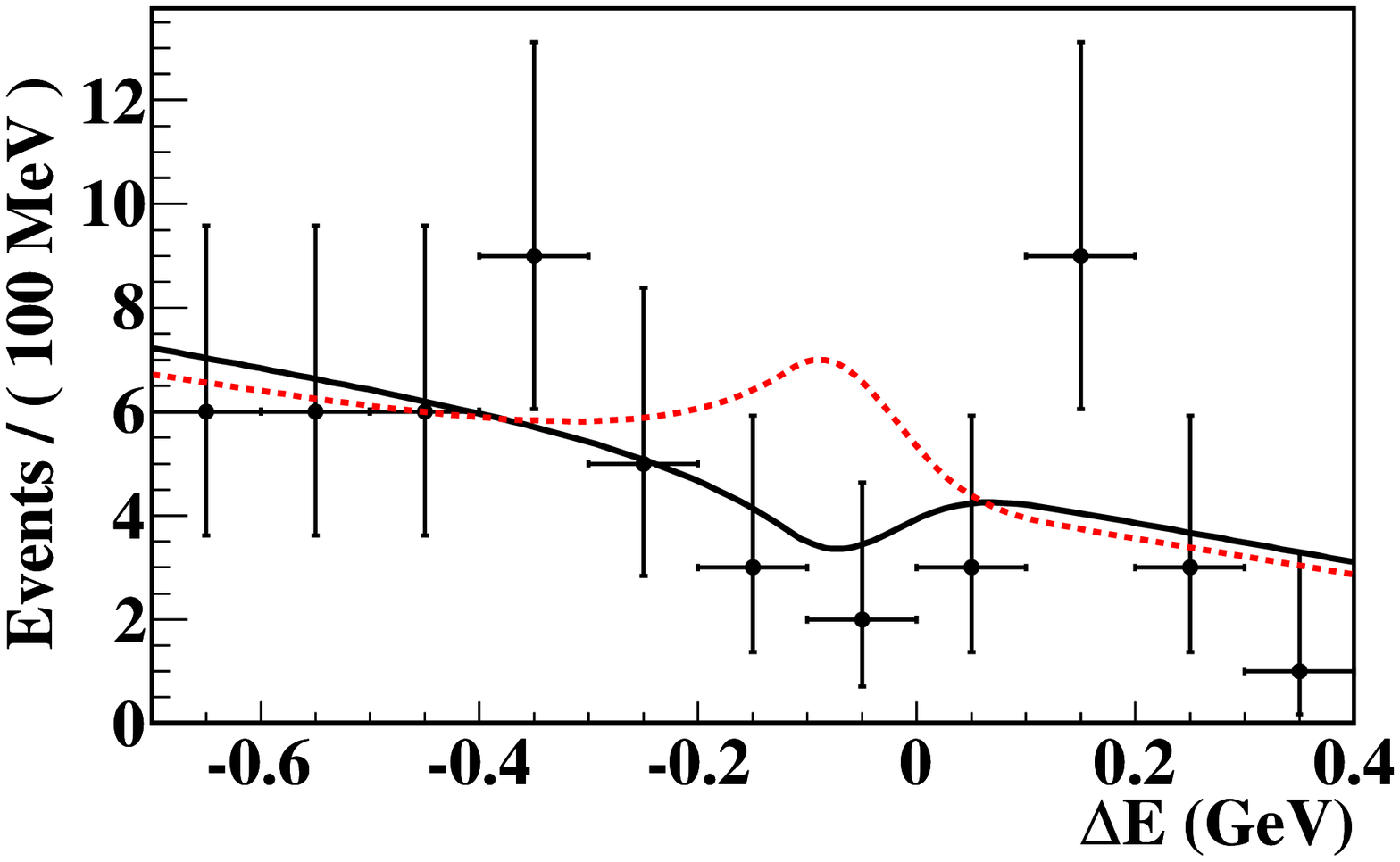}\\
\end{tabular}
\caption{Data fits for the $\BSGG$ analysis. The projections are shown only for events inside the $\BSTBST$ signal region except for the plotted variable. The $\BSTBST$ signal region is defined as $M_{\rm bc} >$ 5.4 GeV/$c^{2}$ and $-$0.3 $ < \Delta E <$ 0.05 GeV. The points with error bars represent the data, the solid black curve represents the total fit function, the dotted red curve represents the fit with the signal yield constrained to its 90$\%$ CL upper limit.}
\label{fig:fit_gg}
\end{figure}

The branching fraction for $\BSPG$ is determined with the relation
\noindent
\begin{linenomath}
\begin{equation}
\mathcal{B}(\BSPG) = \frac{N(\BSPG)}{2 f_s \sigma_{b\bar{b}}^{\Upsilon(5S)} {\cal{L}}_\mathrm{int}\;\epsilon\;\mathcal{B}(\phi\rightarrow K^{+}K^{-})},
\end{equation}
\end{linenomath}
where $N(\BSPG)$ is the signal yield of $\BSPG$, $f_{s}$ is the fraction of $B_{s}^{(*)}B_{s}^{(*)}$ events in the $b\bar{b}$ sample, $\sigma_{b\bar{b}}^{\Upsilon(5S)}$ is the $b\bar{b}$ production cross-section, ${\cal{L}}_{\mathrm{int}}$ is the integrated luminosity at the $\Upsilon(5S)$ energy and $\epsilon$ is the signal selection efficiency. We measure the $\BSPG$ BF to be $(3.6 \pm 0.5 \pm 0.3 \pm 0.6)\times 10^{-5}$, where the first uncertainty is statistical, the second is systematic and the third is due to the uncertainty in $f_{s}$. No statistically significant signal is observed for the decay $\BSGG$ and we measure the single-event sensitivity to be $0.5\times10^{-6}$. We use a Bayesian approach and integrate the likelihood curve from 0 to 90$\%$ of the total integral under the curve to obtain a 90$\%$ CL upper limit of $3.1\times10^{-6}$ on the $\BSGG$ branching fraction. The results are summarised in Table~\ref{table:results}, while the fit results projected onto the signal regions are shown in Figs.~\ref{fig:fit_pg} and ~\ref{fig:fit_gg}.

\begin{table}[!ht]
\centering 
\caption{Results of the $\BSPG$ and $\BSGG$ analyses. The uncertainty in the efficiency calculation is due to the limited MC statistics.}
\vspace{2mm}
\begin{tabular}{p{1.3cm}p{4.3cm}p{2.2cm}}
\hline\hline
                                             & \textbf{$\BSPG$}          & \textbf{$\BSGG$}  \\[0.5ex]
\hline
$\epsilon\;(\%)$                        &   36.1 $\pm$ 0.1           & 14.0 $\pm$ 0.1        \\
$N$                                     &   $91^{+14}_{-13}$         & $-3.9^{+3.7}_{-2.6}$ \\
$\mathcal{B}(10^{-6})$                  &   $36 \pm 5(\mathrm{\small{stat.}}) \pm 3(\mathrm{\small{syst.}}) \pm 6(\small{f_{s}})$       & $<$ 3.1 (90$\%$ CL)              \\
\hline\hline
\end{tabular}
\label{table:results}
\end{table}

The systematic uncertainties summarized in Table~\ref{table:systematic_uncertainty} are associated with the photon reconstruction efficiency, kaon identification efficiency, tracking efficiency, the requirement on $\cal{C}_{\mathrm{NB}}$ that is estimated by comparing the efficiencies in data and MC simulations with the $\BKG$ control sample, limited MC statistics, integrated luminosity, $\sigma_{b\bar{b}}^{\Upsilon(5S)}$, $f_{s}$, PDF parameterization and fit bias. The uncertainty due to PDF parameterization is estimated by the variation in the signal yield when varying each fixed parameter by $\pm 1 \sigma$. To investigate the extent of a bias in the fit, pseudo-experiments are generated using the same PDFs as in the final fit but with the signal and background yields fixed to the expected values. Events generated from the pseudo-experiments are then fitted to obtain the yield and residual distributions. The observed biases of $-0.28 \pm 0.08$ and $-0.10 \pm 0.07$ for $\BSPG$ and $\BSGG$ are corrected and their uncertainties are assigned as systematic uncertainties. The uncertainties due to kaon identification and tracking efficiency are 1.3$\%$ and 0.3$\%$, measured using control samples of $D^{*+}\rightarrow D^{0}\pi^{+}_{\rm{slow}} \rightarrow K^{-}\pi^{+}\pi^{+}_{\rm{slow}}$ and $D^{*+}\rightarrow D^{0}\pi^{+}, D^{0}\rightarrow K_{S}\pi^{+}\pi^{-}, K_{S} \rightarrow \pi^{+} \pi^{-}$ decays, respectively. The uncertainty in the $\phi\rightarrow K^{+}K^{-}$ BF represents another source of systematic uncertainty in the $\BSPG$ analysis, which is taken from~\cite{PDG}.

\begin{table}[!ht]
\centering
\caption{Summary of systematic uncertainties.}
\vspace{2mm}
\begin{tabular}{ccc}
\hline\hline
\multicolumn{3}{c}{Additive systematic uncertainties (events)} \\
\hline
Source                                    & $\BSPG$      & $\BSGG$     \\ [0.5ex]
\hline
PDF parameterization                                &  \fontsize{9.5}{9.5}$^{+1.6}_{-1.7}$ &     $\pm 0.4$            \\ [1mm]
Fit bias                                            &  $\pm 0.1$                           &     $\pm 0.1$            \\
\hline
Total (quadratic sum)                      &  \fontsize{9.5}{9.5}$^{+1.6}_{-1.7}$   &   $\pm 0.4$ \\[1mm]
\hline\hline
\multicolumn{3}{c}{Multiplicative systematic uncertainties ($\%$)} \\
\hline
Source             &$\BSPG$     & $\BSGG$  \\ [0.5ex]
\hline
Photon reconstruction efficiency      &  2.2              &      2 $\times$ 2.2      \\
Kaon identification efficiency        &  2.6              &        -              \\
Tracking efficiency                   &  0.7              &        -              \\
$\cal{C}_{\mathrm{NB}}$ requirement     &  4.8              &      8.7        \\
MC statistics                         &  0.2              &      0.4                   \\
\vspace{0.5mm}
$\mathcal{B}(\phi\rightarrow K^{+}K^{-})$       &  1.0       & -                 \\ [1mm]
$\cal{L}_{\mathrm{int}}$                                          &  \multicolumn{2}{c}{1.3}                \\
$\sigma^{\Upsilon(5S)}_{b\bar{b}}$            &  \multicolumn{2}{c}{4.7}                \\
$f_{s}$                                            &  \multicolumn{2}{c}{17.4}               \\
\hline
Total (quadratic sum)                                     &  19.1       &    20.6           \\
\hline\hline                                                                                            \\
\end{tabular}
\label{table:systematic_uncertainty}
\end{table}


To conclude, we have used the entire Belle $\Upsilon(5S)$ dataset to measure $\mathcal{B}(\BSPG) = (3.6 \pm 0.5(\mathrm{stat.}) \pm 0.3(\mathrm{syst.}) \pm 0.6(f_{s})) \times 10^{-5}$. This improved measurement supersedes our earlier result~\cite{Jean_paper} and is consistent with theoretical predictions~\cite{Greub_Bspg,Ball_Bspg} and a recent LHCb result~\cite{LHCb}. We search for the decay $\BSGG$, where we observe no statistically significant signal, and set the 90$\%$ CL upper limit on its BF at $3.1\times10^{-6}$. This result is an improvement by a factor of about three over the previous published result, consistent with the expected sensitivity for our data sample. This result rules out large contributions to the $\BSGG$ branching fraction from RPV SUSY. It also indicates that the decay $\BSGG$ could be observed at the upcoming Belle \rom{2} experiment with a dedicated run at the $\Upsilon(5S)$ resonance.

We thank the KEKB group for excellent operation of the accelerator; the KEK cryogenics group for efficient solenoid operations; and the KEK computer group, the NII, and PNNL/EMSL for valuable computing and SINET4 network support. We acknowledge support from MEXT, JSPS and Nagoya's TLPRC (Japan); ARC and DIISR (Australia); FWF (Austria); NSFC (China); MSMT (Czechia); CZF, DFG, and VS (Germany); DST (India); INFN (Italy); MEST, NRF, GSDC of KISTI, and WCU (Korea); MNiSW and NCN (Poland); MES and RFAAE (Russia); ARRS (Slovenia); IKERBASQUE and UPV/EHU (Spain); SNSF (Switzerland); NSC and MOE (Taiwan); and DOE and NSF (USA).


\end{document}

%% file: author_pub435.tex
\noaffiliation
\affiliation{University of the Basque Country UPV/EHU, 48080 Bilbao}
\affiliation{Budker Institute of Nuclear Physics SB RAS and Novosibirsk State University, Novosibirsk 630090}
\affiliation{Faculty of Mathematics and Physics, Charles University, 121 16 Prague}
\affiliation{University of Cincinnati, Cincinnati, Ohio 45221}
\affiliation{Deutsches Elektronen--Synchrotron, 22607 Hamburg}
\affiliation{Justus-Liebig-Universit\"at Gie\ss{}en, 35392 Gie\ss{}en}
\affiliation{Gifu University, Gifu 501-1193}
\affiliation{The Graduate University for Advanced Studies, Hayama 240-0193}
\affiliation{Hanyang University, Seoul 133-791}
\affiliation{University of Hawaii, Honolulu, Hawaii 96822}
\affiliation{High Energy Accelerator Research Organization (KEK), Tsukuba 305-0801}
\affiliation{IKERBASQUE, Basque Foundation for Science, 48011 Bilbao}
\affiliation{Indian Institute of Technology Guwahati, Assam 781039}
\affiliation{Indian Institute of Technology Madras, Chennai 600036}
\affiliation{Indiana University, Bloomington, Indiana 47408}
\affiliation{Institute of High Energy Physics, Chinese Academy of Sciences, Beijing 100049}
\affiliation{Institute of High Energy Physics, Vienna 1050}
\affiliation{Institute for High Energy Physics, Protvino 142281}
\affiliation{INFN - Sezione di Torino, 10125 Torino}
\affiliation{Institute for Theoretical and Experimental Physics, Moscow 117218}
\affiliation{J. Stefan Institute, 1000 Ljubljana}
\affiliation{Kanagawa University, Yokohama 221-8686}
\affiliation{Institut f\"ur Experimentelle Kernphysik, Karlsruher Institut f\"ur Technologie, 76131 Karlsruhe}
\affiliation{Kennesaw State University, Kennesaw GA 30144}
\affiliation{Department of Physics, Faculty of Science, King Abdulaziz University, Jeddah 21589}
\affiliation{Korea Institute of Science and Technology Information, Daejeon 305-806}
\affiliation{Korea University, Seoul 136-713}
\affiliation{Kyungpook National University, Daegu 702-701}
\affiliation{\'Ecole Polytechnique F\'ed\'erale de Lausanne (EPFL), Lausanne 1015}
\affiliation{Faculty of Mathematics and Physics, University of Ljubljana, 1000 Ljubljana}
\affiliation{Luther College, Decorah, Iowa 52101}
\affiliation{University of Maribor, 2000 Maribor}
\affiliation{Max-Planck-Institut f\"ur Physik, 80805 M\"unchen}
\affiliation{Moscow Physical Engineering Institute, Moscow 115409}
\affiliation{Graduate School of Science, Nagoya University, Nagoya 464-8602}
\affiliation{Kobayashi-Maskawa Institute, Nagoya University, Nagoya 464-8602}
\affiliation{Nara Women's University, Nara 630-8506}
\affiliation{National Central University, Chung-li 32054}
\affiliation{National United University, Miao Li 36003}
\affiliation{Department of Physics, National Taiwan University, Taipei 10617}
\affiliation{Niigata University, Niigata 950-2181}
\affiliation{Osaka City University, Osaka 558-8585}
\affiliation{Pacific Northwest National Laboratory, Richland, Washington 99352}
\affiliation{Peking University, Beijing 100871}
\affiliation{University of Pittsburgh, Pittsburgh, Pennsylvania 15260}
\affiliation{University of Science and Technology of China, Hefei 230026}
\affiliation{Soongsil University, Seoul 156-743}
\affiliation{Sungkyunkwan University, Suwon 440-746}
\affiliation{School of Physics, University of Sydney, NSW 2006}
\affiliation{Department of Physics, Faculty of Science, University of Tabuk, Tabuk 71451}
\affiliation{Tata Institute of Fundamental Research, Mumbai 400005}
\affiliation{Excellence Cluster Universe, Technische Universit\"at M\"unchen, 85748 Garching}
\affiliation{Toho University, Funabashi 274-8510}
\affiliation{Tohoku University, Sendai 980-8578}
\affiliation{Department of Physics, University of Tokyo, Tokyo 113-0033}
\affiliation{Tokyo Institute of Technology, Tokyo 152-8550}
\affiliation{Tokyo Metropolitan University, Tokyo 192-0397}
\affiliation{CNP, Virginia Polytechnic Institute and State University, Blacksburg, Virginia 24061}
\affiliation{Wayne State University, Detroit, Michigan 48202}
\affiliation{Yonsei University, Seoul 120-749}

\author{D.~Dutta}\affiliation{Indian Institute of Technology Guwahati, Assam 781039} 
\author{B.~Bhuyan}\affiliation{Indian Institute of Technology Guwahati, Assam 781039} 
  \author{A.~Abdesselam}\affiliation{Department of Physics, Faculty of Science, University of Tabuk, Tabuk 71451} 
  \author{I.~Adachi}\affiliation{High Energy Accelerator Research Organization (KEK), Tsukuba 305-0801}\affiliation{The Graduate University for Advanced Studies, Hayama 240-0193} 
  \author{H.~Aihara}\affiliation{Department of Physics, University of Tokyo, Tokyo 113-0033} 
  \author{S.~Al~Said}\affiliation{Department of Physics, Faculty of Science, University of Tabuk, Tabuk 71451}\affiliation{Department of Physics, Faculty of Science, King Abdulaziz University, Jeddah 21589} 
  \author{K.~Arinstein}\affiliation{Budker Institute of Nuclear Physics SB RAS and Novosibirsk State University, Novosibirsk 630090} 
  \author{D.~M.~Asner}\affiliation{Pacific Northwest National Laboratory, Richland, Washington 99352} 
  \author{V.~Aulchenko}\affiliation{Budker Institute of Nuclear Physics SB RAS and Novosibirsk State University, Novosibirsk 630090} 
 \author{T.~Aushev}\affiliation{Moscow Institute of Physics and Technology, Moscow Region 141700}\affiliation{Institute for Theoretical and Experimental Physics, Moscow 117218} 
% ...
\collaboration{The Belle Collaboration}